\documentclass[aps,pra,reprint,superscriptaddress,nofootinbib]{revtex4-1}  

\usepackage{amsfonts} 
\usepackage{amssymb}
\usepackage{graphicx}
\usepackage{tabularx}
\usepackage{tabulary}
\usepackage{upgreek}
\usepackage{subfigure}
\usepackage{psfrag}
\usepackage{lipsum}
\usepackage{color}
\usepackage{braket}
\usepackage{lineno}
\usepackage{mathtools}

\newcommand{\A}{\mathcal{A}}
\newcommand{\B}{\mathcal{B}}
\newcommand{\E}{\mathcal{E}}
\newcommand{\Eeff}{\mathcal{E}_\mathrm{eff}}
\newcommand{\vecEeff}{\vec{\mathcal{E}}_\mathrm{eff}}
\newcommand{\N}{\mathcal{N}}
\newcommand{\R}{\mathcal{R}}
\newcommand{\Nt}{\tilde{\mathcal{N}}}

\newcommand{\wNE}{\omega^{\N \E}}

\newcommand{\upperlimitFeldman}{$1.1\times 10^{-29}$~$e\cdot \textrm{cm}$}

\begin{document}
\title{Attaining the shot-noise-limit in the ACME measurement of the electron electric dipole moment}

\author{C. D. Panda \noaffiliation}
\affiliation{Department of Physics, Harvard University, Cambridge, Massachusetts 02138, USA}
\author{C. Meisenhelder \noaffiliation}
\affiliation{Department of Physics, Harvard University, Cambridge, Massachusetts 02138, USA}
\author{M. Verma \noaffiliation}
\affiliation{Department of Physics, University of Toronto, Toronto, Ontario M5S 1A7, Canada}
\author{D. G. Ang \noaffiliation}
\affiliation{Department of Physics, Harvard University, Cambridge, Massachusetts 02138, USA}
\author{J. Chow \noaffiliation}
\affiliation{Department of Physics, Yale University, New Haven, Connecticut 06511, USA}
\author{Z. Lasner \noaffiliation}
\affiliation{Department of Physics, Yale University, New Haven, Connecticut 06511, USA}
\author{X. Wu \noaffiliation}
\affiliation{Department of Physics, Harvard University, Cambridge, Massachusetts 02138, USA}
\affiliation{Department of Physics, Yale University, New Haven, Connecticut 06511, USA}
\author{D. DeMille \noaffiliation}
\affiliation{Department of Physics, Yale University, New Haven, Connecticut 06511, USA}
\author{J. M. Doyle \noaffiliation}
\affiliation{Department of Physics, Harvard University, Cambridge, Massachusetts 02138, USA}
\author{G. Gabrielse \noaffiliation}
\affiliation{Center for Fundamental Physics, Northwestern University, Evanston, Illinois 60208, USA}

\begin{abstract}
Experimental searches for the electron electric dipole moment, $d_e$, probe new physics beyond the Standard Model.  Recently, the ACME Collaboration set a new limit of $|d_e| <$ \upperlimitFeldman  ~[Nature \textbf{562}, 355 (2018)], constraining time reversal symmetry (T) violating physics in the 3-100 TeV energy scale. ACME extracts $d_e$ from the measurement of electron spin precession due to the thorium monoxide (ThO) molecule's internal electric field. This recent ACME II measurement achieved an order of magnitude increased sensitivity over ACME I by reducing both statistical and systematic uncertainties in the measurement of the electric dipole precession frequency. The ACME II statistical uncertainty was a factor of 1.7 above the ideal shot-noise limit. We have since traced this excess noise to timing imperfections. When the experimental imperfections are eliminated, we show that shot noise limit is attained by acquiring noise-free data in the same configuration as ACME II.

\end{abstract}

\maketitle

\section*{Introduction}

The electric dipole moment of the electron, $\vec{d}_e$, is an asymmetric charge distribution along the particle’s spin, $\vec{s}$. Theories of physics beyond the Standard Model often include new particles with masses of 3-100 TeV/$c^2$ whose interaction with the electron include T-violating phases and lead to $d_e \approx 10^{-27}-10^{-30}~e \cdot \mathrm{cm}$  \cite{Cesarotti2018,Fuyuto2018,Nakai2017,Barr1993,Pospelov2005,Engel2013,Bernreuther1991}, orders of magnitude higher than the value predicted by the Standard Model \cite{Pospelov1991, Pospelov2014}. Measurements of $d_e$ with increased precision probe for new physics in this energy range \cite{ACMECollaboration2018}.

Recent advances in the measurement of $d_e$ \cite{ACMECollaboration2018, Baron2014,ACMECollaboration2016,Cairncross2017,Kara2012} have relied on the exceptionally high internal effective electric field $\Eeff$ of heavy polar molecules. We perform our measurement in the H$^3 \Delta_1$ state of ThO, which provides $\Eeff=78~\mathrm{GV/cm}$ \cite{Denis2016, Skripnikov2016}. In the presence of $d_e$, this gives rise to an energy shift $U=-\vec{d}_e \cdot \vecEeff$. 

We measure this energy shift $U$ by observing electron spin precession in parallel uniform applied electric ($\vec{\E}=\E \hat{z}$) and magnetic fields ($\vec{\B}=\B \hat{z}$). We control the spin of the H$^3 \Delta_1$ molecular state, $\vec{S}$, which is proportional to the spin of the electron $\vec{s}$. To initialize the measurement, we use a linearly polarized laser propagating along $\hat{z}$, the axis of the applied fields $\vec{\E}$ and $\vec{\B}$, to align $\vec{S}$ along the fixed direction given by the polarization of the laser light \cite{Panda2016,ACMECollaboration2018}. The $\vec{S}$ vector is in the $xy$ plane and perpendicular to $\hat{z}$.

We allow $\vec{S}$ to precess under the torques of the applied magnetic field $\vec{\B}$ and $\vecEeff$ on the magnetic and electric dipole moments associated with $\vec{S}$. We measure the precession angle $\phi=\omega \tau$, where the precession frequency is
\begin{equation}
\omega  \approx \frac{-\mu \tilde{\mathcal{B}}\left|\mathcal{B}_{z}\right|-\tilde{\mathcal{N}}\tilde{\mathcal{E}}d_{e}\mathcal{E}_{\mathrm{eff}}}{\hbar}, \label{eq:phase}
\end{equation}
and $\tau$  is the spin precession time, $|\mathcal{B}_z| = |\vec{\mathcal{B}}\cdot\hat{z}|$, $\tilde{\mathcal{B}}=\mathrm{sgn}(\vec{\mathcal{B}}\cdot\hat{z})$, $\tilde{\mathcal{E}}=\mathrm{sgn}(\vec{\mathcal{E}}\cdot\hat{z})$, and $\mu=\mu_\mathrm{B}g_H$, where $g_H= -0.0044$ is the $g$-factor of the H, J=1 state \cite{Kirilov2013} and $\mu_\mathrm{B}$ is the Bohr magneton. We extract the precession time, $\tau$, from the change of $\phi$ that comes from reversing the applied magnetic field, $\phi^{\mathcal{B}}=-\mu|\mathcal{B}_z|\tau/\hbar$. We then compute the angular precession frequency, $\omega=\phi/\tau$.

We use pairs of states in the H $^3\Delta_1$ manifold that correspond to $\vecEeff$ being aligned and anti-aligned with the applied $\vec{\E}$, labeled by the quantum number $\Nt=\mathrm{sgn}(\vecEeff \cdot \vec{\E})$ \cite{ACMECollaboration2016}. These states are spectroscopically resolved, and tuning our lasers to be resonant with either $\Nt=\pm1$ allows us to reverse the direction of $\vecEeff$ independently of the direction of $\vec{\E}$. To extract the contribution of $d_e$ to $\omega$, we reverse the direction of $\vecEeff$ either by reversing the laboratory field $\vec{\E}$ or by changing the state $\Nt=\pm1$ used in the measurement. By denoting this contribution as $\wNE$, we obtain $d_e=-\hbar \wNE/\Eeff$.

The standard quantum limit for the uncertainty in the measurement of $d_e$ is determined by shot noise: that is, for $N$ detected molecules, $\delta_{d_e}^{s-n}=(2 \tau \Eeff\sqrt{N})^{-1}$ \cite{ACMECollaboration2016}. However, technical noise sources can make $\delta_{d_e} >\delta_{d_e}^{s-n}$ \cite{Kirilov2013}. Unfortunately, a previously unidentified source produced a form of technical noise that increased the ACME II statistical uncertainty in the measurement of $d_e$ by a factor of 1.7 above shot noise. In this work we trace this excess noise to imperfect hardware timing. We verify that the excess noise was accounted for appropriately in the ACME II analysis. We also show that with the timing imperfections under control, the shot noise limit can be attained. Eliminating this error will allow future ACME measurements to obtain higher sensitivity.




\section*{Measurement of the precession frequency through fast polarization switching}

\begin{figure}
\begin{center}
\includegraphics[width=0.5\textwidth]
{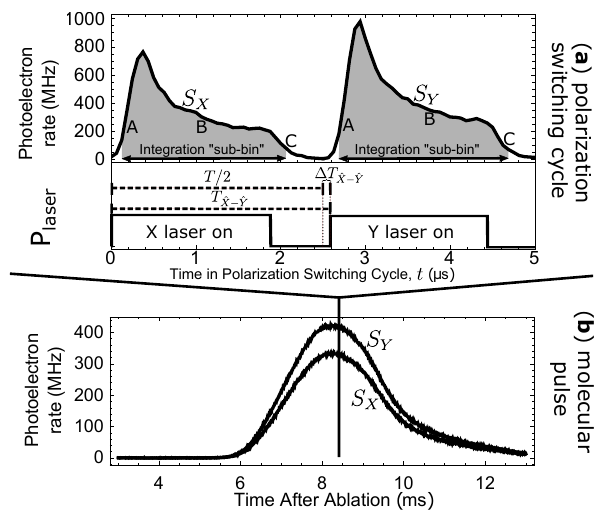}
\caption{
 \textbf{Switching timescales}.
\textbf{(a)} Fluorescence signal size vs. time in an $\hat{X}$, $\hat{Y}$ polarization cycle. The integration ``sub-bin" typically used in ACME II is shown in gray. $T_{\hat{X}-\hat{Y}}$ is the time delay between the $\hat{X}$ and $\hat{Y}$ laser pulses. $\Delta T_{\hat{X}-\hat{Y}}=T_{\hat{X}-\hat{Y}}-T/2$ is the asymmetric relative time delay between the optical $\hat{X}$, $\hat{Y}$ pulses. A, B, and C denote regions of distinct quantum state population dynamics.
\textbf{(b)} Measured molecular fluorescence signal trace (25 pulses averaged) vs. time. Shown signals are averaged over the entire $\hat{X}$ and $\hat{Y}$ polarization cycles from \textbf{(a)}.
\label{fig:switches}}
\end{center}
\end{figure}

We measure the precession frequency $\omega$ by exciting the H–I transition with laser light (703 nm) linearly polarized along direction $\hat{\epsilon}$. This yields fluorescence signals with intensity $S_{\hat{\epsilon}}$, which depends on the angle between $\hat{\epsilon}$ and $\vec{S}$. To remove excess technical noise due to fluctuations in molecule number, we excite the molecules with two alternating orthogonal linear polarizations, $\hat{\epsilon}=\hat{X},\hat{Y}$, by modulating $\hat{\epsilon}$ sufficiently rapidly (period 5~$\mu$s) so that each molecule is addressed by both polarizations as it passes through the laser beam \cite{ACMECollaboration2018, ACMECollaboration2016}. We record the corresponding fluorescence signals $S_X(t)$ and $S_Y(t)$ from the decay of I to the ground state X (wavelength 512 nm; see Fig. \ref{fig:switches}), as a function of time within the polarization switching cycle, $t$. 

In ACME II, fluorescence was recorded using a data acquisition (DAQ) digitizer\footnote{NI PXI-5171R FPGA.} operating at a sampling rate of 16 MSa/s. At this sampling rate, each acquired sample contained signal integrated over $T_{\rm dig}=62.5$~ns. Each polarization switching cycle (period $T=5$~$\mu$s) contained 80 samples, with the first (last) 40 assigned to signals with polarization $\hat{X}$ ($\hat{Y}$). We labeled the digitized signals at each point as $S_X^i$ ($S_Y^i$), where $i\in\{ 1,2,3...40 \}$ labels the digitization point starting at time $t_i = (i-1) T_{\rm dig}$. The first point in the polarization cycle, $S_X^1$, was chosen consistently throughout the analysis as the point where the $\hat{X}$ laser turns on, i. e. the first point where $S_X>0$. We computed integrated fluorescence signals by summing over samples within a chosen region of time between when a given polarization is turned on and when the next polarization is turned on, $sb$, which we referred to as an integration ``sub-bin'' (see Fig. \ref{fig:switches})
\begin{equation} 
S_{X}=\sum \limits_{i \in sb} S_{X}^i~(S_{Y}=\sum \limits_{i \in sb} S_{Y}^i)
\end{equation} 
The ``sub-bin" is common to both $\hat{X}$ and $\hat{Y}$ polarization cycles. Typically, we used $sb=\{3,4,5...34\}$ in ACME II. Typical ACME II integrated fluorescence signals plotted as a function of time after ablation are shown in Figure \ref{fig:switches}(b).

We then compute the asymmetry
\begin{equation}
\A=\frac{S_X-S_Y}{S_X+S_Y}=\mathcal{C} \cos[2(\phi-\theta)], \label{eq:asym}
\end{equation}
where the contrast $\mathcal{C}$ is 95\% $\pm$ 2\% on average and $\hat{X}$ is defined to be at an angle $\theta$ with respect to the horizontal lab axis $\hat{x}$ in the $x–y$ plane. We measure $\mathcal{C}$ by dithering $\theta$ between two nearby values, $\tilde{\theta}=\pm1$, that differ by 0.2 rad (11.5 degrees).  We then compute the precession frequency, 
\begin{equation}
\omega=\frac{\phi}{\tau}=\frac{\mathcal{A}}{2 \mathcal{C} \tau},
\end{equation}
from the asymmetry, $\A$, contrast, $\mathcal{C}$, and precession time, $\tau$.

To implement the fast polarization switching scheme experimentally, we overlap two laser beams with orthogonal $\hat{X}$ and $\hat{Y}$ polarizations, which we switch alternatively on and off rapidly using acousto-optical modulators (AOMs) \cite{ACMECollaboration2016, Vutha2010}. The two beams are combined on a polarizing beamsplitter which rejects any possible polarization imperfections. In ACME II, the $\hat{X}$ and $\hat{Y}$ pulses each had a duration of 1.9~$\mu$s, with a nominal 0.6~$\mu$s delay between them. Given the 115 ns lifetime of the I state (Fig. \ref{fig:switches}), this delay was sufficient to reduce the overlap between fluorescence signals arising from excitation by the different laser polarizations.

The shape of the time-modulated fluorescence signal, $S(t)$, is given by the quantum state population dynamics resulting from the properties of the readout molecular states H and I, and the laser beam intensity spatial and time profiles. Immediately after the laser is switched on, there is a rapid increase in fluorescence as molecules in the laser beam are quickly excited (region A in Fig. \ref{fig:switches}). When $\Omega_r t \ll  1$, where $\Omega_r \sim 2 \pi \times 3~\mathrm{MHz}$ is the Rabi frequency of the readout H-I transition, the fluorescence magnitude increases as $S(t) \propto \Omega_r^2 t^2$ . Later, when $\Omega_r t \ge 1$, population is roughly evenly mixed between the H and I states, causing $S(t)$ to decay nearly exponentially with a time constant of $2 \tau_I$ (region B in Fig. \ref{fig:switches}), where $\tau_I\approx 115~\mathrm{ns}$ is the lifetime of the I state. During the time the laser is on, molecules continually enter the laser beam, such that the nearly exponential decay approaches a constant fluorescence rate in the steady state. After the laser turns off, the signal decays exponentially with time constant $\tau_I$ (region C in Fig. \ref{fig:switches}).

One important parameter in this polarization switching scheme is the time delay between the $\hat{X}$ and $\hat{Y}$ laser pulses, $T_{\hat{X}-\hat{Y}}$ (Fig. \ref{fig:switches}). Ideally, $T_{\hat{X}-\hat{Y}}=T/2$, where $T$ is the polarization switching period ($T=5~\mu$s in ACME II). However, since the laser intensity modulation is performed by AOMs, there is an additional delay in the timing of the $\hat{X}$, $\hat{Y}$ optical pulses relative to the electronic trigger pulses due to the propagation time of the acoustic wave in the acousto-optic crystal \cite{Degenhardt2005,Falke2012}. This propagation delay is sensitive to the alignment and spatial intensity profile of the laser beam and the geometry of the specific AOM crystal used. We found that in our apparatus, it could vary due to manual realignment of the laser beam through the AOM by up to 200 ns. Such alignment was typically done every several days during ACME II. 

During ACME II, we corrected for this additional relative delay between the $\hat{X},\hat{Y}$ optical pulses by manually adding in the experiment timing structure a time delay between the $\hat{X},\hat{Y}$ electronic trigger pulses, such that the asymmetric relative time delay between the optical $\hat{X}$, $\hat{Y}$ pulses, $\Delta T_{\hat{X}-\hat{Y}}=T_{\hat{X}-\hat{Y}}-T/2$, is minimized, i.e. $\Delta T_{\hat{X}-\hat{Y}}\approx0$. This was implemented during ACME II by observing both $\hat{X}$ and $\hat{Y}$ optical pulses on the same photodiode and matching the optical dead-times between the signals. Using this technique, we could set $\Delta T_{\hat{X}-\hat{Y}}=0$ with $\sim 40$~ns precision, better than the timing corresponding to one digitizer sample (62.5 ns). However, we show below that even this imprecision in setting $\Delta T_{\hat{X}-\hat{Y}}=0$ is important; with any nonzero residual value, the asymmetry acquires a dependence on time within the polarization switching cycle that can cause frequency noise when combined with technical timing noise.

\section*{Frequency noise in ACME II}

During the ACME II measurement sequence, we performed a set of 7 binary switches of experimental parameters\footnote{The switches in a superblock are described in \cite{ACMECollaboration2018}, but the details are unimportant here.} that allowed us to compute the frequency component due to $d_e$, $\omega^{\N \E}$. The time scales of the switches ranged from the fastest (0.6 s) to slowest (10 minutes). Each set of $2^7$ states ($\sim 20$ minutes acquisition time), corresponding to the 7 switches, represented a ``superblock''. 

During data acquisition, we averaged 25 molecular pulses together to form a ``trace" (0.5 s averaging time). Within a trace, we computed $\mathcal{A}$ for each polarization cycle. We then averaged 20 consecutive cycles into a single ``group," with uncertainty defined as the standard error in the mean of the group. The uncertainties of each group were consistent with the level due to shot noise on our photoelectron signals. We then used standard uncertainty propagation to compute uncertainties from an entire superblock. 

The ACME II dataset consisted of $\sim1000$ superblocks acquired over a period of 2 months. The majority of the data was consistent with a distribution nearly Gaussian near its center, but with an excess of points in the tails \cite{ACMECollaboration2018}. In addition, the scatter in the superblock data was found to be larger than expected from group-level uncertainties. The excess noise was present equally in all $2^7$ states of the experiment.  Furthermore, the relative magnitude of the noise with respect to shot-noise did not vary as a function of time within the molecular pulse. The excess noise in the precession frequency had one contribution that was proportional to the $\B$-field magnitude, and another that was independent of $\B$. We discuss the two separately.

We quantify the magnitude of the noise by computing the reduced chi-square per degree of freedom of the dataset, $\chi_r^2$. The $\B-$field dependent component of the excess noise increased the scatter of the ACME II superblock data to $\chi_r^2 \sim 7$, for data acquired at the largest applied $\B$-field magnitude, $|\mathcal{B}_z|=26$ mG. As described previously \cite{ACMECollaboration2018}, we reduced this noise contribution by acquiring most data at lower magnetic field magnitudes, $|\mathcal{B}_z|\in\{0.7,1.3,2.6\}~\mathrm{mG}$, where the associated increase in $\chi^2_r$ is negligible.


We focus the discussion in this paper on the $\B$-independent component of excess noise, which limited the sensitivity of ACME II. The statistical uncertainty was $\sim1.7$ times larger than that expected from shot-noise, corresponding to a reduced chi-squared statistic of the superblock data of $\chi_r^2 \sim 3$. 

\section*{Diagnosis of excess noise sources}

To characterize this excess noise source, we perform a noise diagnosis in an experimental setup that is similar to ACME II, but without actually executing any of the 7 binary switches. Furthermore, we perform our analysis on data from single molecular pulses, rather than averaging 25 consecutive pulses in a ``trace'', as was done in ACME II. This allows us to observe the properties of our measurement directly at fast timescales, before switching and averaging obscure important underlying characteristics of our measurement that we ultimately found were leading to frequency noise.


\subsection*{Mechanism causing variable trigger-to-digitizer delays}

Using this diagnosis method, we found that one ingredient that causes frequency noise is variation in the triggering of the acquisition of the individual molecular pulses. Such variation can occur in our system due to a lack of synchronization between the signals triggering the polarization switching AOMs and those triggering the DAQ digitizer. In ACME II, a common high precision timing and delay generator\footnote{SRS DG645.} provided TTL pulses that acted as triggers for the RF switches that modulated the polarization switching AOMs on and off. The same pulse generator acted as a trigger for the digitization sequence. These two trigger signals were phase locked to suppress their relative timing jitter to $<25~\mathrm{ps}$. 

However, during this diagnosis of noise sources, we find that the synchronization of the laser polarization pulses with the DAQ digitization events was not, in fact, consistent with the low jitter between these trigger pulses. The reason is that our particular DAQ digitizer uses an internal clock to perform the timing of the sampling and digitization process, rather than responding directly to an external trigger. Hence, if the timing generator and the DAQ internal clock are not explicitly synchronized, there is an uncontrolled delay between the DAQ acquisition trigger pulse and the actual start of digitization. 

During ACME II, this asynchrony caused molecular pulses to have their digitization begin with varying time delays relative to the AOM triggers controlling the $\hat{X}-\hat{Y}$ polarization switch, with magnitude of up to $\sim100$ ns. Each subsequent molecular pulse (triggered at a rate of 50 Hz) was deterministically offset from the previous by $\sim 10$ ns. This timing offset is consistent with  the inaccuracy ($5\times 10^{-7}$) of the internal clock of the DAQ device. When the delay reached $\sim100$~ns (every 10 molecular pulses), it reset to 0, creating a periodic sawtooth pattern.

For the current tests, we eliminate this varying trigger-to-digitizer delay by using an external clock\footnote{10 MHz from a Rubidium reference clock, SRS FS725.} to synchronize the electronic signals triggering the polarization switching AOMs with the internal clock of the DAQ digitizer. However, proper synchronization of the DAQ digitizer to an external clock also required a firmware update of the digitizer. This originally made the noise difficult to identify. Furthermore, low-jitter synchronization ($<25$ ps) between experiment timing and actual digitization events is only possible when the digitizer's sampling rate is set to be an integer divisor of the 250 MSa/s internal clock rate. This was not the case in ACME II, where the digitizer sampling rate was set to 16 MSa/s.

To achieve minimum timing variation between digitization events and polarization switching, we implement a modified timing structure from that used in ACME II. We choose a 12.5 MSa/s digitization rate in these tests, commensurate with the digitizer's internal clock rate. (Faster rates are not possible in the current setup due to the limited data transfer rate of the computer system performing the acquisition.) At this sampling rate, each acquired sample contains signal integrated over 80 ns (compared to 62.5 ns in ACME II). To ensure an even number of digitization samples in a polarization switching cycle, we set the polarization switching frequency to 250 kHz (compared to 200 kHz in ACME II), such that each full polarization cycle contains exactly 50 digitizer samples, 25 corresponding to the $\hat{X}$ and $\hat{Y}$ polarization bins, respectively. The dead-time between the $\hat{X}$ and $\hat{Y}$ halves of the polarization cycle, when both laser polarizations are off, is set to 0.8 $\mu$s (0.6 $\mu$s in ACME II).


\subsection*{Asymmetry dependence on $\Delta T_{\hat{X}-\hat{Y}}$}

\begin{figure}[tb]
\includegraphics[scale=0.6]{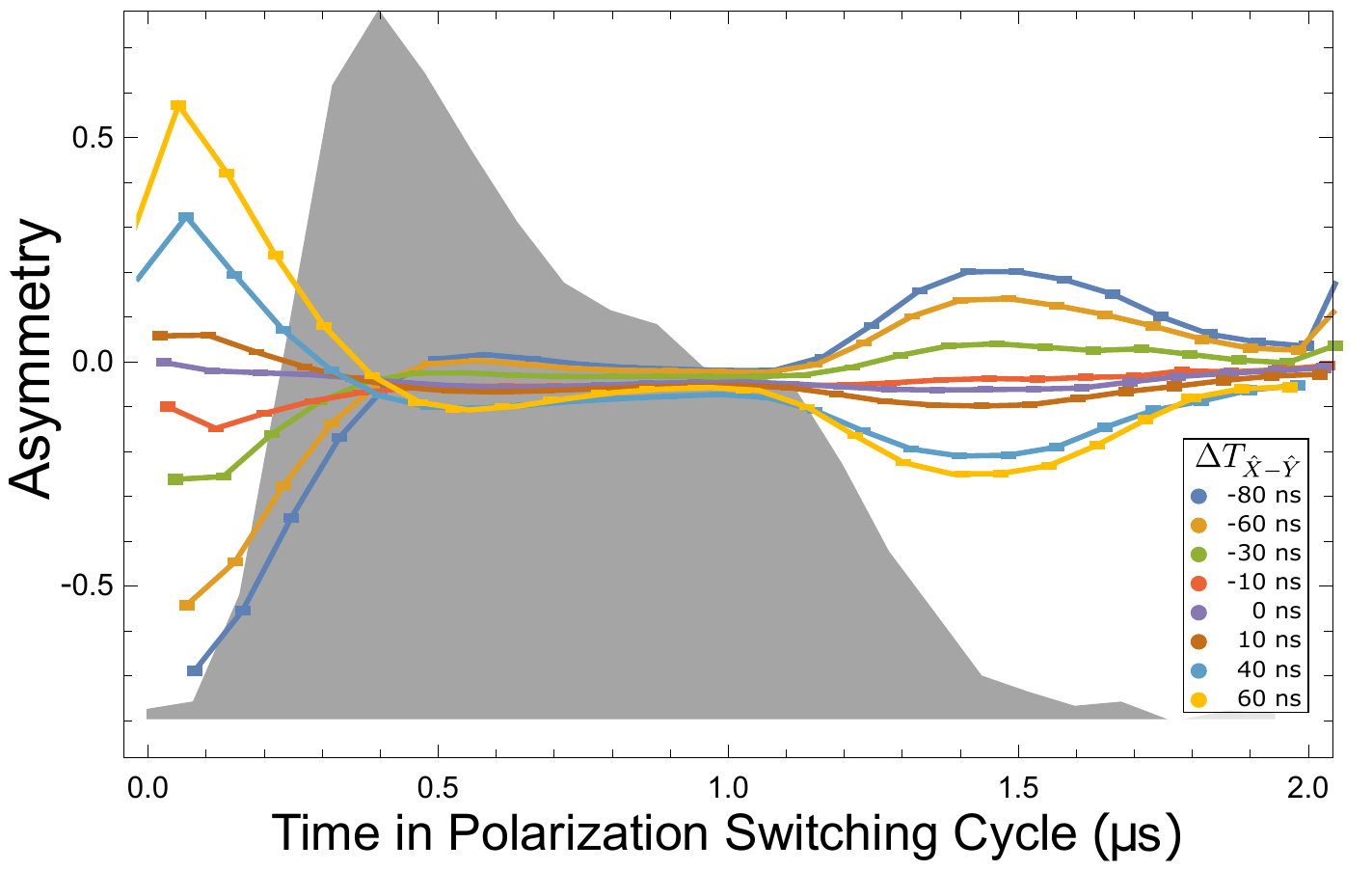}
\centering
\caption{\textbf{Asymmetry, $\mathcal{A} (t)$, versus time in polarization switching cycle, for various values of $\Delta T_{\hat{X}-\hat{Y}}$.} The average signal, $S(t)$, is shown in gray. The magnitude of $\mathcal{A} (t)$ at any given time is proportional to the asymmetric $\hat{X}$, $\hat{Y}$ pulse time delay parameter, $\Delta T_{\hat{X}-\hat{Y}}$,  and to the slope of the fluorescence signal in the polarization bin, $dS(t)/dt$. The asymmetries shown here are calculated by averaging 200 consecutive molecular pulses (4 seconds averaging time).
\label{fig:Plasymshapes}}
\end{figure}

As described above, for ACME II, we typically computed a``time-averaged" asymmetry by averaging the digitizer samples in the chosen integration ``sub-bin," defined in the same way for both $\hat{X}$ and $\hat{Y}$ halves of the polarization cycle. In contrast, we analyze the data used for the noise tests described here by calculating the asymmetry for each single acquired digitizer sample in the $\hat{X}$ and $\hat{Y}$ bins: 
\begin{equation}
\mathcal{A}^i=\frac{S_X^i-S_Y^i}{S_X^i+S_Y^i},
\end{equation}
where now $i\in\{1,2,3...25\}$ due to the new full polarization switching period ($T=4$~$\mu$s) and new digitization rate (12.5 MSa/s). This results in asymmetry values which we use to show the dependence of the noise and asymmetry offset on the time in the polarization switching cycle.

Figure \ref{fig:Plasymshapes} shows the asymmetry, $\mathcal{A}^i=\mathcal{A}(t_i)$, as a function of time in the polarization switching cycle.  We observe that the asymmetric relative time delay between the optical $\hat{X}$, $\hat{Y}$ pulses, $\Delta T_{\hat{X}-\hat{Y}}$ (see Fig. \ref{fig:switches}),  has a large effect on the shape and magnitude of the resulting asymmetry and its dependence on time within the polarization cycle. This occurs because when $\Delta T_{\hat{X}-\hat{Y}}\neq 0$, there is a difference in the acquisition times of $S_X(t_i)$ and $S_Y(t_i)$ relative to when the laser light with that polarization is switched on. This causes the computed asymmetry within the polarization switching cycle, $\mathcal{A}(t_i)$, which should nominally be constant, to have a time-dependent difference from its mean, $\Delta\mathcal{A}(t)$. When $\Delta T_{\hat{X}-\hat{Y}} \ll T$, we can approximate this variation in the asymmetry as
\begin{equation} \label{eq:noise}
\Delta \mathcal{A}(t) \approx \frac{1}{2 S (t)} \frac{d S(t)}{d t} \Delta T_{\hat{X}-\hat{Y}},
\end{equation}
where $S(t)$ is the signal averaged over the $\hat{X}$ and $\hat{Y}$ polarizations, $S(t)=(S_{X}(t)+S_{Y}(t))/2$.

We note that $\Delta \mathcal{A} (t)$ is independent of any of the experiment switches performed routinely as a part of the ACME II superblock structure. Therefore, in the channels of interest in the experiment, all of which are odd under at least one of these switches, offsets due to $\Delta\mathcal{A}(t)$ are cancelled. In particular, we have searched for and not observed any systematic variation of the $\omega^{\N \E}$ frequency, or any of the other odd frequency channels, correlated with time within the polarization switching cycle \cite{Panda2018}. In addition, the $\tilde{\mathcal{P}}$ and $\tilde{\R}$ switches (described in detail in \cite{ACMECollaboration2016}) each interchange the roles of the $\hat{X}$ and $\hat{Y}$ readout laser beams \cite{ACMECollaboration2016,ACMECollaboration2018}, reversing the sign of the asymmetry: $\mathcal{A}(t) \xrightarrow {\tilde{\mathcal{P}},\tilde{\R}} -\mathcal{A}(t)$. This transformation subtracts $\Delta \mathcal{A} (t)$ in the $\tilde{\mathcal{P}}$, $\tilde{\R}$-even channels (such as the $\wNE$ channel, which is used to compute $d_e$), so that on average the presence of $\Delta \mathcal{A}(t)$ cannot systematically shift the measurement of $\phi$ and $\omega$.

\subsection*{Technical variable trigger-to-digitizer delays lead to asymmetry noise}

\begin{figure}[tbp]
\includegraphics[scale=0.8]{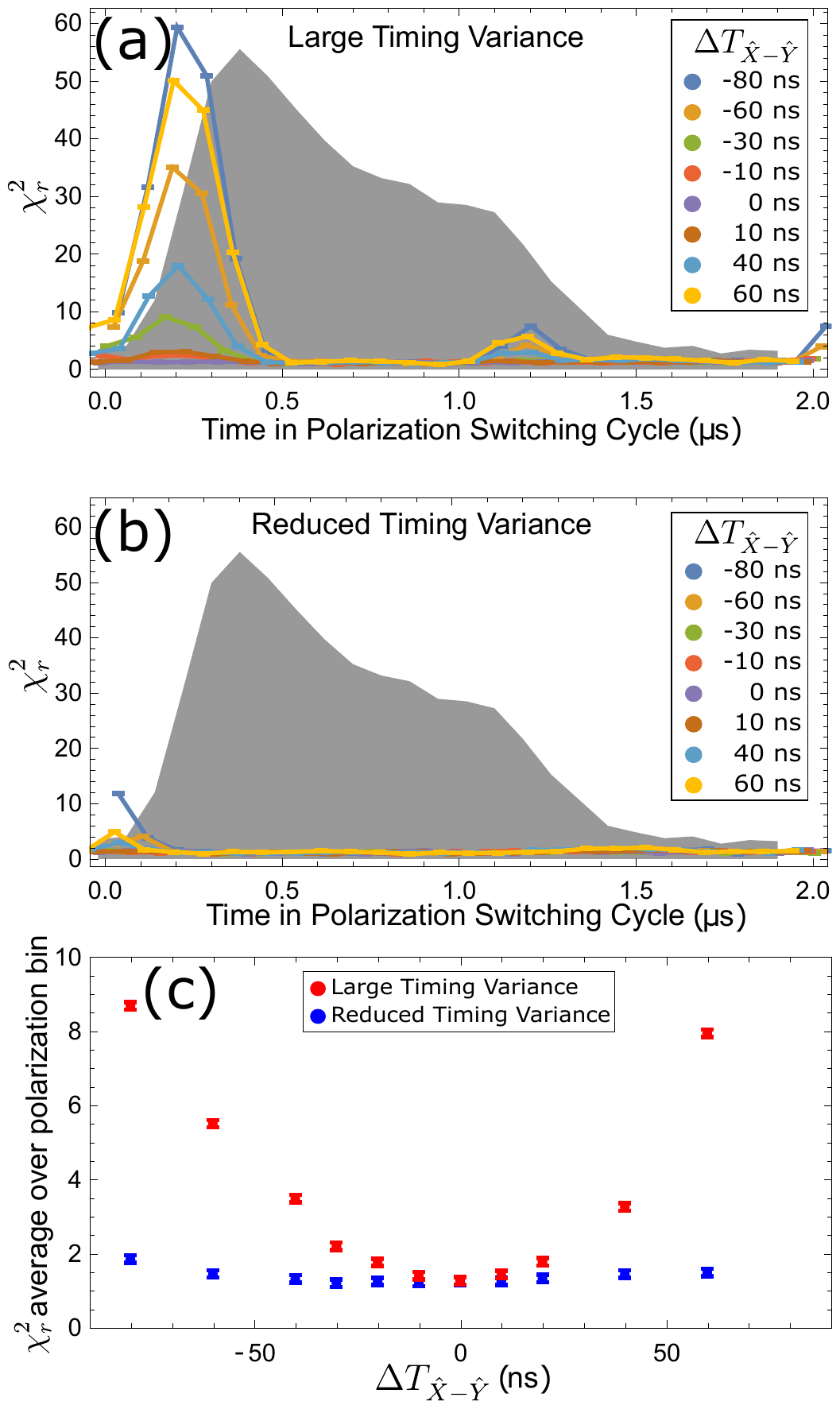}
\centering
\caption{\textbf{Variation of asymmetry noise within the polarization bin with $\Delta T_{\hat{X}-\hat{Y}}$. } Measured magnitude of excess noise, parameterized by $\chi^2_r$, versus time within the polarization switching cycle, for various values of $\Delta T_{\hat{X}-\hat{Y}}$, \textbf{(a)} with large timing noise, and \textbf{(b)} when timing noise is reduced. The noise is larger where the slope of the fluorescence signal ($S(t)$ shown in gray), $d S(t)/dt$, is larger. \textbf{(c)} $\chi^2_r$ averaged over the entire time in the polarization switching cycle, shown as a function of $\Delta T_{\hat{X}-\hat{Y}}$, for large timing variance and when timing variance is reduced. All $\chi^2_r$ values are calculated for 200 consecutive molecular pulses, acquired over 4 s.
\label{fig:Plnoisesyncnosync}}
\end{figure}
This dependence of asymmetry on time within the polarization switching cycle can, however, cause noise in the asymmetry. This arises when such a non-zero $\Delta \mathcal{A}(t)$ is present together with a variation in the trigger-to-digitizer delay relative to the start of the polarization switching laser pulses. This noise appears not only in the raw asymmetry, but also leaks into the channels which are odd with respect to the performed switches, if the variation of the trigger timing takes place on time scales shorter than the fastest experimental switch defining any such channel. For example, any $\Nt$-odd switch parity signal will exhibit this noise if the trigger timing varies on timescales that are faster than the $\Nt$ experiment switch (every 0.6~s). This is shown in Figures \ref{fig:Plnoisesyncnosync}(a) and \ref{fig:Plnoisesyncnosync}(b), when there is a large amount of timing variance (as present during the ACME II dataset) or with reduced technical timing variance, respectively. We achieve the two configurations by either making commensurate or not the DAQ internal clock and the external clock that defines the polarization switching times, as described above. When the clocks are not commensurate, the noise also propagates equally into all computed odd and even switch channels since the technical timing variation timescale ($\approx$10 pulses $=$ 0.2~s) is faster than the timescale of the fastest experiment switch $\Nt$ (0.6~s).

In the presence of large timing variation, the computed $\chi^2_r$ for the set of 200 molecular pulses is largest at the beginning and end of the polarization switching optical pulse. The data is consistent with our model (Eq. \ref{eq:noise}), where noise is proportional to the time dependent asymmetry shown in Figure \ref{fig:Plasymshapes}, $\chi^2_r \propto \Delta \mathcal{A} (t)^2 \propto (\frac{d S(t)}{dt} \Delta T_{\hat{X}-\hat{Y}})^2$. Figure \ref{fig:Plnoisesyncnosync}(c) shows $\chi^2_r$ averaged over the entire polarization cycle (integration ``sub-bin" $sb = \{1,2,...,25\}$) as a function of $\Delta T_{\hat{X}-\hat{Y}}$, in the presence of timing variance and with timing variance reduced. This demonstrates the reduced magnitude of noise with lower timing variance and when  $\Delta T_{\hat{X}-\hat{Y}}=0$.

\section*{Control and suppression of noise}

With the noise mechanism understood, we reduce the magnitude of the excess noise by suppressing the experiment imperfections that contribute to it. As shown in Figure \ref{fig:Plnoisesyncnosync}(c), we can reduce the noise by suppressing DAQ timing variance and/or setting $\Delta T_{\hat{X}-\hat{Y}}$ to zero. Each of these parameters can be suppressed by several orders of magnitude compared to values that were typical in ACME II. Since the suppression is multiplicative, this source of noise can be greatly reduced in future ACME experiments.

\begin{figure}[t]
\includegraphics[scale=0.55]{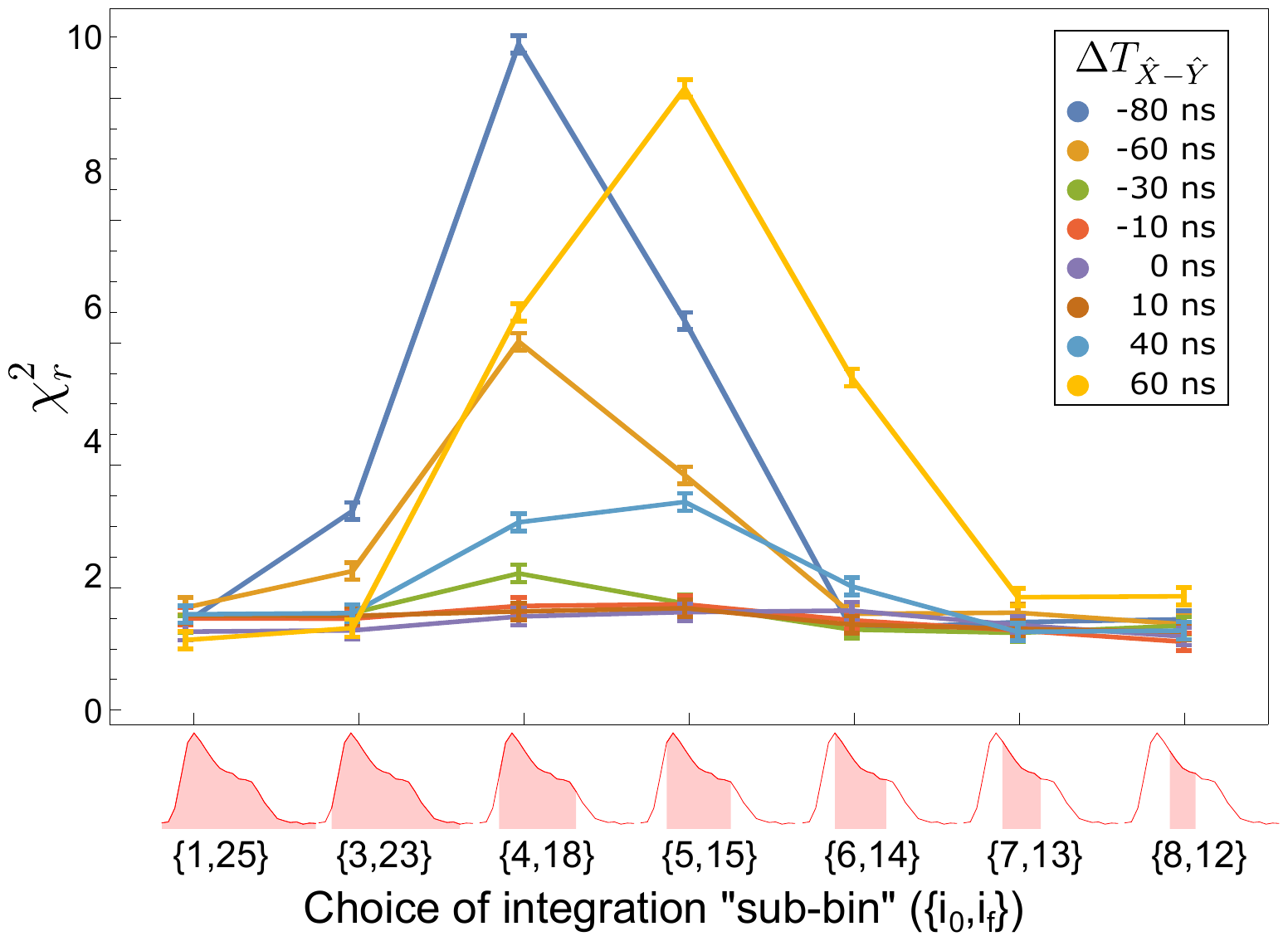}
\centering
\caption{\textbf{Dependence of the excess noise on choice of integration sub-bin.} The noise ($\chi^2_r$) is reduced when the integration sub-bin is chosen such that it does not begin or end with the samples with largest amount of excess noise. This behavior is consistent for all values of $\Delta T_{\hat{X}-\hat{Y}}$. The shown data is acquired in the ``large timing noise'' configuration, where the clocks of the timing box and DAQ digitizer are not synchronized. All $\chi^2_r$ values are calculated for 200 consecutive molecular pulses, acquired over 4~s.
\label{fig:Plnoisesubbins}}
\end{figure}

A third method could also be used to further suppress this source of frequency noise if necessary. As shown in Eq. \ref{eq:noise}, $\Delta \mathcal{A}(t)$ is proportional to the slope of the signal, $d S/ d t$. When summed over the entire ``sub-bin", as typically done during ACME II data analysis, $\Delta \mathcal{A}$ is only a function of signals at the beginning and end of the ``sub-bin" integration time: $\Delta \mathcal{A} = \int_{t_{i0}}^{t_{if}} \Delta \mathcal{A} (t) dt$, where $i0$ and $if$ are the first and last indexes in the integration ''sub-bin". There is no dependence of $\Delta \mathcal{A}$ on the intermediary points, i.e. for $i0<i<if$. This means that the noise can be minimized if we choose the ``sub-bin" such that variation in signal at both the start and end times, $S^{i0}$, $S^{if}$ are minimized. This behavior is shown in Figure \ref{fig:Plnoisesubbins}.

Finally, we verified the suppression of the asymmetry noise when using all three methods simultaneously (minimized $\Delta T_{\hat{X}-\hat{Y}}$, reduced timing variance, using a full ``sub-bin" of $sb=\{1,2,...,25\}$). Under these conditions, we acquired 12 superblocks of data with the same sets of switches and parameters as in the ACME II experiment. This produced data consistent with a Gaussian distribution out to its tails with $\chi^2_r=0.87\pm0.40$, consistent with 1. This confirmed the suppression of this (and any other) sources of noise to below the ACME II shot-noise limited statistical uncertainty.

A further alternative method of suppressing such noise in the future is by performing one of the experimental switches that changes the sign of the signals of interest at a timescale that is faster than that of any timing variation. This can be achieved, for example, by  using the $\tilde{\mathcal{P}}$ switch, which is currently implemented using AOMs and could be performed at a faster timescale. Another option is performing the $\Nt$ switch at faster timescales. We could, for example, switch $\Nt$ every molecular pulse, 25 times faster than in ACME II.

\section*{Conclusion}
We have understood and quantified a technical timing variation mechanism that accounts for the excess noise present in the ACME II measurement of $d_e$. The mechanism which added unbiased noise the measurement of the asymmetry was due to a combination of timing variance between the DAQ digitizer and polarization switching events, and an asymmetric relative time delay between the $\hat{X}$ and $\hat{Y}$ polarization switching optical pulses. Such noise mechanisms can be a concern for experiments that, like ACME, compare different experimental states by rapidly switching between them.

We showed here that the noise can be suppressed by reducing the two timing imperfections that contribute to it and by integrating over a larger sub-bin within the polarization switching signal. We verified suppression to a level below the ACME II experiment shot-noise limited sensitivity, by acquiring noise-free data in the same configuration as ACME II. The noise reduction represents a factor of 1.7 increase in the statistical sensitivity of future ACME experiments, compared to ACME II, for this effect alone. Based on the model for its origin, we expect that this source of technical noise is suppressed by several orders of magnitude below its ACME II level.

\section*{Acknowledgements}
This work was performed as part of the ACME Collaboration, to whom we are grateful for its contributions,
and was supported by the NSF.

\bibliography{references}

\end{document}